\documentstyle[12pt]{article}

\input epsf

\begingroup\makeatletter\ifx\SetFigFont\undefined
\def\x#1#2#3#4#5#6#7\relax{\def\x{#1#2#3#4#5#6}}%
\expandafter\x\fmtname xxxxxx\relax \def\y{splain}%
\ifx\x\y   
\gdef\SetFigFont#1#2#3{%
  \ifnum #1<17\tiny\else \ifnum #1<20\small\else
  \ifnum #1<24\normalsize\else \ifnum #1<29\large\else
  \ifnum #1<34\Large\else \ifnum #1<41\LARGE\else
     \huge\fi\fi\fi\fi\fi\fi
  \csname #3\endcsname}%
\else
\gdef\SetFigFont#1#2#3{\begingroup
  \count@#1\relax \ifnum 25<\count@\count@25\fi
  \def\x{\endgroup\@setsize\SetFigFont{#2pt}}%
  \expandafter\x
    \csname \romannumeral\the\count@ pt\expandafter\endcsname
    \csname @\romannumeral\the\count@ pt\endcsname
  \csname #3\endcsname}%
\fi
\fi\endgroup

\advance \topmargin by -\headheight
\advance \topmargin by -\headsep

\evensidemargin \oddsidemargin
\vfil
\begin{document}

\begin{flushright}
SLAC--PUB--7565\\
hep-ph/9706530  \\
June 1997
\end{flushright}
\bigskip\bigskip

\thispagestyle{empty}
\flushbottom
\vfill
\centerline{\large\bf PION PRODUCTION FROM BAKED-ALASKA}
\vspace*{0.035truein}
\centerline{{\large\bf DISORIENTED CHIRAL CONDENSATE}
    \footnote{\baselineskip=14pt
Work supported in part by funds provided by the Foundation Blanceflor
Boncompagni-Ludovisi,
\hfill\break \hbox{\hskip.2in}
P.P.A.R.C., U. S. Department of Energy contract
DE-AC03-76SF00\-515, the American Trust for
\hfill\break \hbox{\hskip.2in} 
Oxford University (George
Eastman Visiting Professorship), CSN and KV (Sweden), ORS and 
\hfill\break \hbox{\hskip.2in}
OOB (Oxford), and the Sir Richard Stapley Educational Trust (Kent).}}

\vspace{22pt}

  \centerline{\bf G. Amelino-Camelia, J. D. Bjorken
  \footnote{\baselineskip=14pt Permanent address:
  Stanford Linear Accelerator Center, Stanford University, Stanford,
  \hfill \break \hbox{\hskip.2in} California 94309}, 
  and S. E. Larsson}
\vspace{8pt}
  \centerline{\it Theoretical Physics, University of Oxford}
  \centerline{\it 1 Keble Road, Oxford OX1 3NP}
  \centerline{\it United Kingdom}
\vspace*{0.9cm}

\begin{abstract}

We study the various stages of the evolution of chiral condensates
disoriented via the ``baked-alaska'' mechanism, in which the
condensates are described as the products of external sources localized
on the light cone. Our analysis is based on the classical equations of
motion of either the linear or the nonlinear sigma model. We use the
associated framework of coherent states and, especially, their source
functions to make the connection to the distribution functions for the
produced particles. We also compare our classical approach with a
mean-field calculation which includes a certain class of quantum
corrections.

\end{abstract}
\vfill

\centerline{Submitted to Physical Review D.}
\vfill
\baselineskip 12pt plus 0.2pt minus 0.2pt

\newpage

\section{Introduction}

Recently, in order to interpret events with a deficit or excess of
neutral pions observed in cosmic ray experiments, there has been
increased interest in the conjecture that it might be possible to
produce in high-energy collisions {\it disoriented chiral condensates}
(DCCs), {\it i.e.} correlated regions of space-time wherein the quark
condensate, $\langle 0| q_L  \bar{q}_R |0 \rangle$, is chirally rotated
from its usual orientation in isospin space.

On the theoretical side there has been great interest (see, {\it e.g.},
Refs. \cite{anselm}-\cite{CKMP}) both in the development of technical
tools suitable for the description of this possible phenomenon, and in
the exploration of the possibilities opened by DCCs as probes of the
structure of quantum chromodynamics, most notably in relation to the
chiral phase transition.

The idea that such DCCs might be produced in high energy collisions at
existing or planned hadron or heavy-ion accelerators has generated some
experimental interest. In particular, one of us is co-spokesman for a
Fermilab experiment \cite{t864} looking for DCCs in hadron-hadron
collisions. In high energy $p$-${\bar p}$ collisions which lead to a
sizable multiplicity of produced particles, but not necessarily with
high-$p_T$ jets in the final state, the time evolution is
quasi-macroscopic, because the hadronization time can be rather large,
up to 3-5 $fm$. At times $t$ before hadronization, the initial state
partons, produced in a volume much smaller than a cubic fermi, may
stream outward at essentially the speed of light in all directions,
occupying the surface of a sphere of radius $t$ (in units such that the
speed of light is 1). Most of the outgoing energy/momentum is expected
to be concentrated near the light cone, {\it i.e.} on the surface of
the expanding `fireball'.

However, the interior of the fireball is also an interesting place. If
its energy density is low enough, the interior should look very similar
to the vacuum, with an associated non-vanishing quark condensate. Since
the energy density from the intrinsic chiral symmetry breaking is small
\cite{pisa,kota}, and since the fireball surface isolates the interior
from the exterior of the light cone, it is reasonable to consider the
possibility that well inside the light cone the quark condensate might
be chirally rotated from its usual orientation. At late times this
disoriented vacuum would relax back to ordinary vacuum, via radiation
of its collective modes, the pions. The properties of the radiated
pions should be strongly affected by the semiclassical, coherent nature
of the process. In particular, one may expect anomalously large
event-by-event fluctuations in the ratio of the number of charged pions
to neutral pions produced. Assuming that the event-by-event deviation
of the quark condensate from its usual orientation be random, one finds
\cite{anselm,bj,krishna,blaizot,kota,karma} that the distribution
$P(f)$ of the neutral fraction 
\begin{eqnarray} 
f \equiv {N_{\pi^0} \over N_{\pi^0} + N_{\pi^+} + N_{\pi^-} } \equiv
{N_{\pi^0} \over N_{tot} } ~,
\label{fdef} 
\end{eqnarray} 
is given by
\begin{eqnarray} 
{dP \over df} = {1 \over 2 \, \sqrt{f}} ~,
\label{radice} 
\end{eqnarray}
at large $N_{tot}$. Most notably, this implies that for ``DCC pions''
the probability of finding extreme values of $f$ is very different from
ordinary pion production (which is given by a binomial distribution),
in which the fluctuations are expected to be peaked at $f \! = \! 1/3$
and fall exponentially away from the peak.

Experimental DCC searches \cite{t864} are thus far largely based on the
structure of Eq.~(\ref{radice}). However, it is probable that this
robust property will not be sufficient on its own for the
identification of phenomena involving DCCs. In particular, it appears
that an understanding of the geometry of the phenomena would be very
useful for experimental searches.

In this paper we develop a description of ``baked-alaska'' (hot
fireball surface with cooler and disoriented inside) DCCs which is well
suited for the study of realistic geometries. Our analysis is based on
the classical equations of motion of either the linear or the nonlinear
sigma model and we use the associated framework of coherent states to
make the connection to the distribution functions for the produced
particles.

In the next section we review the coherent-state formalism with
emphasis on the associated particle flux, which is seen as directly
related to classical sources localized on the light cone. Sec. 3 is
devoted to the sigma model description of ``baked-alaska DCCs'',
defined via the above-mentioned mechanism involving a hot fireball
shell and its cooler (and sometimes disoriented) interior. In Sec. 4 we
derive a simple solution of the classical nonlinear sigma model, and
follow the evolution of the associated coherent state from beginning to
end, including a derivation of the flux of pions. In Sec. 5 we discuss
how our analysis can be generalized to the linear sigma model, and
establish connections with a somewhat similar model studied by Lampert,
Dawson, and Cooper \cite{lampert}. Sec. 6 is devoted to closing remarks.

\section{Particle production
from classical sources} \label{classapp}

\subsection{Coherent state formalism} \label{coherentstates}

In the following sections we will be investigating a problem which in
the classical limit is described by the field equation 
\begin{eqnarray}
        (\Box + m^2)\phi(x) & = & J(x) ~,
\label{boxeq}
\end{eqnarray}
for appropriate boundary conditions. The properties of the quanta
associated with a field, such as $\phi$ in (\ref{boxeq}), that are
produced by a classical current source $J$ can be studied using the
{\it coherent state} defined by
\begin{eqnarray}
|J\rangle & \equiv & \exp \left[-{1 \over 2}\int d\tilde k~
|\tilde J(k)|^2\right] \exp \left[\int d\tilde  k~
\tilde J(k)\hat a^{\dagger}(k)\right] |0\rangle
~, \label{cohegene}
\end{eqnarray}
where $\tilde J(k)$ is the Fourier transform of the source $J(x)$,
$\hat a^\dagger$ is the usual creation operator, 
and the integral measure is given by
\begin{eqnarray}
        d\tilde k & \equiv & {d^4k \over (2\pi)^4}~2\pi
\delta(k^2-m^2) \theta(k^0) 
~. \label{meas}
\end{eqnarray}
Note that, because of the integration measure, the actual contribution
of $\tilde J(k)$ to the integrals in (\ref{cohegene}) comes from the
mass-shell with $k^0 \! > \! 0$, and sometimes it proves convenient to
introduce the notation
\begin{eqnarray}
        \tilde J(\vec k) & \equiv & \tilde J(k)|_{k^0=\omega_k}
~, \label{etamasshell}
\end{eqnarray}
where
\begin{eqnarray}
        \omega_k & \equiv & \sqrt{{\vec k}^2 + m^2} \geq 0
~. \label{omegak}
\end{eqnarray}

We remind the reader that the relation between the coherent state
$|J\rangle$ and the source $J$ can be derived from the familiar
(text-book) analysis of the $S$-matrix associated with $J$. Provided
that $J$ is sufficiently well localized in space and time (so that the
idea of a scattering process is well defined), this $S$-matrix is given
by
\begin{eqnarray}
        \hat S & = & :\exp \int d^4y~\hat\phi_f(y)J(y):
~, \label{smatbefore}
\end{eqnarray}
where $\hat\phi_f(y)$ is the free scalar field operator. Taking into
account the normal ordering, one can rewrite $\hat S$ as
\begin{eqnarray}
        \hat S & = &  \exp \left[{1 \over 2}\int d\tilde k~
|\tilde J(k)|^2\right] \exp \left[-i \int
d^4y~J(y)\hat\phi_f^-(y)\right] \exp \left[-i \int
d^4y~J(y)\hat\phi_f^+(y)\right] \nonumber \\ 
               & = &  \exp \left[{1 \over 2}\int d\tilde k~
|\tilde J(k)|^2\right] \exp \left[ -i \int d\tilde k~ \tilde J(k) \hat
a^{\dagger}(k)\right] \exp \left[ -i \int d\tilde k~ \tilde J(k) \hat
a(k)\right]
~. \label{smatafter}
\end{eqnarray}
In particular, this implies that the ``out'' state $\hat
S^{-1}|0\rangle$ corresponding to a scattering process having the
vacuum $|0\rangle$ as ``in'' state, is given by the coherent state of
Eq.~(\ref{cohegene}).

The most appropriate tool for the description of the particle
production associated with the ``out'' state is the generating
functional, which in the present case takes the form
\begin{eqnarray}
        G[z_\phi] & = & \exp \left[ \int d\tilde k~|\tilde J(k)|^2 
                (z_\phi(k) - 1) \right]~, \label{generfirst}
\end{eqnarray}
from which one can obtain all the inclusive and exclusive factorial
moments. In particular, the inclusive spectrum of particles as a
function of momentum is given by
\begin{eqnarray}
2\omega_k {dN_{\phi} \over d^3 \vec k} 
= 2\omega_k \left[ {\delta G[z_\phi] \over \delta z_\phi}
\right]_{z_\phi = 1} 
= {1\over (2\pi)^3}|\tilde J(\vec k)|^2
~. \label{fluxgene}
\end{eqnarray}

Note that all higher correlation functions vanish for a coherent state,
as is easily shown by taking further derivatives of the generating
functional. Thus, we see that the main object of interest in this
description is the function $\tilde J(\vec k)$.

\subsection{Extracting $\tilde J$ from the field}

Since one's intuition is that the flux of particles depends directly on
the field, it is instructive to see how the value of $\tilde J(\vec k)$
can be derived from the (classical) field at late times. Assuming that
the ``in'' state has vanishing classical field (it is the classical
``vacuum'' state), $\phi_{in}(x) = 0$, we can write
\begin{eqnarray}
        \phi_{out}(x) & = & \int d^4y~ D^{(-)}(x-y) \, J(y)
~, \label{phioutofgint}
\end{eqnarray}
where $D^{(-)} \equiv D_{ret} - D_{adv}$ is the difference between the
retarded and advanced Green functions. Taking the Fourier transform
(and using the convolution theorem) we get
\begin{eqnarray}
\tilde \phi_{out}(k) & = & \tilde D^{(-)}(k) \, \tilde J(k) \nonumber\\
& = & 2\pi \, \delta(k^2-m^2) \, \epsilon(k^0) ~ i \, \tilde J(k)
~, \label{phioutofg}
\end{eqnarray}
where we have substituted for the explicit form of $\tilde D^{(-)}(k)$
in the last line. It is then useful to consider the three dimensional
Fourier transform of $\phi_{out}(x)$. We use the following conventions
\begin{eqnarray}
        \tilde f(k) & = & \int d^4x~e^{ikx}f(x) ~, \label{four4}\\
        \tilde f^{(3)}(t,\vec k) & = & \int
d^3x~e^{-i\vec{k}.\vec{x}}f(t,\vec x) 
~, \label{four3}
\end{eqnarray}
which imply
\begin{eqnarray}
        \tilde f^{(3)}(t,\vec k) & = & \int {dk^0 \over (2\pi)}~
e^{-ik^0t}\tilde f(k) ~. \label{3DFT}
\end{eqnarray}
From (\ref{phioutofg}) and (\ref{3DFT}) it follows that
the three dimensional Fourier transform
of $\phi_{out}(x)$ is given by
\begin{eqnarray}
        \tilde \phi^{(3)}_{out}(t,\vec k) & = & \int {dk^0 \over
(2\pi)}~e^{-ik^0t}~2\pi\delta(k^2-m^2)\epsilon(k^0)~i\tilde J(k) \nonumber\\
                                     & = & {1 \over \omega_k}\left\{ [i
\tilde J(\vec k)] e^{-i\omega_k t} +  [i\tilde J(\vec k)]^*
e^{i\omega_k t} \right\}
~, \label{four3out}
\end{eqnarray}
where we have used the fact that $J(x)$ is real
\begin{eqnarray}
        \tilde J(k)^* & = & \tilde J(-k)
~. \label{jreal}
\end{eqnarray}
Upon observing that (\ref{four3out}) also implies 
\begin{eqnarray}
        \dot{\tilde\phi}{\vphantom{\phi}}^{(3)}_{out}(t,\vec k)
& = & \tilde{\dot 
\phi}{\vphantom{\phi}}^{(3)}_{out}(t,\vec k) \nonumber \\
                       & = & -i \left\{ [i
\tilde J(\vec k)] e^{-i\omega_k t} - [i\tilde J(\vec k)]^*
e^{i\omega_k t} \right\}  
~, \label{four3outtimeder}
\end{eqnarray}
we can solve for $\tilde J(\vec k)$ in terms 
of $\tilde \phi^{(3)}_{out}$ and its time derivative
\begin{eqnarray}
        \tilde J(\vec k) & = & 
-i e^{i\omega_k t}\left[ \omega_k \tilde \phi^{(3)}_{out}(t,\vec k) 
+ i\dot{\tilde \phi}{\vphantom{\phi}}^{(3)}_{out}(t,\vec k)\right]  
~. \label{Jfromfield}
\end{eqnarray}
This relation is extremely useful for problems set up in such a way
that what is known is encoded in the equations of motion plus the
initial field configuration; in fact, (\ref{Jfromfield}) allows us to
derive an associated current source (which in turn describes the
particle flux via (\ref{fluxgene})) from the late-time field.

\section{Sigma-model description of baked-alaska DCCs}

The $O(4)$ sigma-model is typically used as a model of hadron dynamics
in DCC studies. It is simple enough to be treatable, has the correct
chiral symmetry properties, and describes the low energy phenomenology
of pions. However it must be kept in mind that it is, at best, a crude
approximation to the chiral effective low-energy Lagrangian of QCD.

The Lagrangian density of the linear sigma-model is
\begin{eqnarray}
{\cal L} = {1 \over 2} (\partial_\mu \sigma)^2
+ {1 \over 2} (\partial_\mu \vec{\pi})^2
- {\lambda \over 4} (\sigma^2+\vec{\pi}^2 - f_\pi^2)^2
+ H \sigma
~.\label{lagrlin}
\end{eqnarray}
$H \! = \! 0$ in the chiral limit, $m_\pi \! = \! 0$.

A meaningful sigma-model description can start at some small proper
time, of order 0.2-0.3~$fm$, near the light cone, when the collective
coordinates $\sigma$ and $\pi$ become relevant\cite{bjminn,gm}. At this
early proper time the distribution of the chiral field
\begin{eqnarray}
\Phi \equiv  \sigma +  i \vec{\pi} \cdot \vec{\tau}
~,\label{udef}
\end{eqnarray}
can be expected to be noisy, but with $\langle\Phi\rangle = 0$.

As proper time increases the field $\Phi$ rolls into a minimum with
$\Phi^\dagger \Phi \! = \! f_\pi^2$, and during this ``rolling phase''
the pion mass can be imaginary, leading to unstable growth of the
Goldstone modes\cite{krishna,bjminn}. Since, as mentioned in the
Introduction, the energy density from the intrinsic chiral symmetry
breaking is small\cite{pisa,kota}, and the fireball surface isolates
the interior from the exterior of the light cone, it is reasonable to
expect that the interior of the light cone ends up in a disoriented
vacuum.

At late times such a region of disoriented vacuum with a given isospin
orientation would relax back to ordinary vacuum $\langle\Phi\rangle =
\langle\sigma\rangle = \! f_\pi$, radiating pions with the same isospin
orientation.

It is also reasonable to expect that approximations based on the
replacement of the full linear sigma-model by the simpler nonlinear
sigma-model, with Lagrangian density given (in the chiral limit) by
\begin{eqnarray} 
{\cal L} = {1 \over 2} (\partial_\mu \sigma)^2 + {1 \over 2}
(\partial_\mu \vec{\pi})^2 ~,~~~ \rm{with} ~ \sigma^2 + \vec{\pi}^2 =
\it{f}_\pi^2 ~, 
\label{lagrnonlin} 
\end{eqnarray} 
could be reliably used at times late enough so that the chiral field
has already rolled into a minimum with $\Phi^\dagger \Phi \! \equiv \!
\sigma^2 + \vec{\pi}^2 \! = \! f_\pi^2$.

In modeling these stages of evolution, the chiral limit can be safely
taken as long as the proper time is small compared to $m_\pi^{-1}$,
while at proper times of order (1 to 2)$m_\pi^{-1}$ the pion mass can
no longer be neglected. At sufficiently large proper times, one should
\cite{bjminn} decompose the DCC field into physical-pion normal modes
and let them propagate out to infinity as free states.

Returning to the coherent state formalism reviewed in
Section~\ref{coherentstates}, we note that in the linear sigma-model we
are dealing with four scalar fields. As a result, the appropriate
``out'' state is a coherent state characterized by four current
densities
\begin{eqnarray}
|\tilde J_\alpha\rangle & = & \exp \left[-{1 \over 2}\int
d\tilde k~ \sum_{\alpha = 0}^3
|\tilde J_\alpha(k)|^2\right] \exp
\left[ i \int d\tilde k~ \sum_{\alpha = 0}^3 \tilde J_\alpha(k)\hat
a^{\dagger}_\alpha(k)\right] |0\rangle
~, \label{cohefourcurr}
\end{eqnarray}
with the corresponding generating functional
\begin{eqnarray}
G[z_0,z_1,z_2,z_3] & = & \exp \left[\int d\tilde k~ \sum_{\alpha =
0}^3 |\tilde J_\alpha(k)|^2(z_\alpha(k) - 1) \right]
~. \label{generfourcurr}
\end{eqnarray}

When not interested in observing $\sigma$ quanta (or working within the
nonlinear sigma-model) one can simply set its fugacity variable $z_0$
to unity. This amounts to setting to zero the source associated with
$\sigma$, which we choose to be $\tilde J_0$. This leaves the
generating functional for the $\pi^1$, $\pi^2$ and $\pi^3$ fields
\begin{eqnarray}
G[z_1,z_2,z_3] & = & \exp \left[ {\int d\tilde k~ \sum_{i =1}^3 
|\tilde J_i(k)|^2(z_i(k) - 1)}\right]
~. \label{cohefourthree}
\end{eqnarray}
The physically observed fields are $\pi^0$, $\pi^+$ and $\pi^-$, which
are associated with the creation operators
\begin{eqnarray}
        {\hat a}^\dagger_0 & = & {\hat a}^\dagger_3 \\
        {\hat a}^\dagger_\pm & 
= & {1 \over \sqrt{2}}({\hat a}^\dagger_1 \pm i{\hat a}^\dagger_2)
~. \label{creatpm}
\end{eqnarray}
Introducing the corresponding source currents\footnote{This $\tilde
J_0$ should not be confused with the source term for $\sigma$,
eliminated in the preceding discussion.}
\begin{eqnarray}
\tilde J_0(k) & = & \tilde J_3(k) \\
\tilde J_\pm(k) & 
= & {1 \over \sqrt{2}} (\tilde J_1(k) \mp i \tilde J_2(k))
~, \label{jpm}
\end{eqnarray}
one finds that
\begin{eqnarray}
\sum_{i=1}^3 \tilde J_i(k){{\hat a}^\dagger_i}(k) & = & \tilde
J_0(k){{\hat a}^\dagger_0}(k) +  \tilde J_+(k){{\hat a}^\dagger_+}(k)
+ \tilde J_-(k){{\hat a}^\dagger_-}(k)
~, \label{joldjpm}
\end{eqnarray}
and the generating functional for the $\pi^0$,
$\pi^+$, and $\pi^-$ fields is given by
\begin{eqnarray}
G[z_0,z_+,z_-] & = & \exp \left[ \int d\tilde k~ \left(
|\tilde J_0(k)|^2(z_0(k) - 1) \right. \right. \\
&   & \left. ~~~~~~\vphantom{\int} \left. + |\tilde J_+(k)|^2(z_+(k) - 1) 
+ |\tilde J_-(k)|^2(z_-(k) - 1) \right) \right]
~. \label{generpm}
\end{eqnarray}

For the DCC picture advocated in the following,
the class of sources 
\begin{eqnarray}
\tilde J_i(k) & = & |\tilde J(k)| n_i
~, \label{goodj}
\end{eqnarray}
where $i \! = \! 1,2,3$ and the $n_i$ are real constants 
such that $\sum_{i=1}^3 n_i^2 \! = \! 1$,
is of particular interest.
From (\ref{goodj}) one clearly has
\begin{eqnarray}
 \tilde J_0(k) & = & |\tilde J(k)| n_3 ~,\nonumber\\
 \tilde J_\pm(k) & = & {|\tilde J(k)| \over \sqrt{2}} (n_1 \mp i n_2)
~, \label{goodjpm}
\end{eqnarray}
and the generating functional takes the form
\begin{eqnarray}
        G[z_0,z_+,z_-] & = & \exp \left[ \int d\tilde k~ 
|\tilde J(k)|^2 \left( (n_3)^2(z_0(k) - 1) \vphantom{{(n_1)^2\over 2}}
\right. \right. \\ 
                       &   & \left. \left.~~~~~~~~ + {(n_1)^2 +
(n_2)^2 \over 2} \left[(z_+(k) - 1) + (z_-(k) - 1)\right] \right) \right]
~. \label{genergoodj}
\end{eqnarray}
By realizing that $f=(n_3)^2$,
with $f$ being the neutral fraction as in the Introduction,
one can rewrite this generating functional as
\begin{eqnarray}
G[z_0,z_+,z_-] & = & \exp \left[ \int d\tilde k~ |\tilde
J(k)|^2 \left( f(z_0(k) - 1) \vphantom{{(1-f) \over 2}}\right. \right. \\
       &   & \left. \left.~~~~~~~~ + {(1-f) \over 2}
\left[(z_+(k) - 1) + (z_-(k) - 1)\right] \right) \right]
~. \label{genergoodjf}
\end{eqnarray}

This generating functional is appropriate for the description of events
with initial conditions parametrized by a given $f$. However, for DCC
production one must average over initial conditions~\cite{apw} with the
appropriate weights $P(f)$. In such cases one can introduce the
following type of generating functional
\begin{eqnarray}
        \bar G[z_0,z_+,z_-] & = & \int_0^1 df~ P(f) \exp \left[ \int
d\tilde k~ |\tilde J(k)|^2 \left( f(z_0(k) - 1) \vphantom{{(1-f) \over
2}} \right. \right. \\
                       &   & \left. \left.~~~~~~~~ + {(1-f) \over 2}
\left[(z_+(k) - 1) + (z_-(k) - 1)\right] \right) \right]
~. \label{generave}
\end{eqnarray}
In particular, assuming that the initial $n_i$ are distributed with
equal probability over the $\sum_{i=1}^3 n_i^2 \! = \! 1$ sphere, we
get the characteristic form 
\begin{eqnarray} 
P(f) & = & {1 \over
2\sqrt{f}} ~ 
\label{pfave} 
\end{eqnarray} 
for the distribution of $f$.

Finally, we observe that the above generating functionals with
independent description of $\pi^+$ and $\pi^-$ production can be turned
into generating functionals for charged particles by fixing
\begin{eqnarray}
        z_{ch} \equiv z_+ = z_-
~. \label{zch}
\end{eqnarray}
For example,
for the generating functional (\ref{generave})
and the $P(f)$ of (\ref{pfave}) one finds
\begin{eqnarray*}
        \bar G[z_0,z_{ch}] & = & \int_0^1 {df \over 2\sqrt{f}}~ 
\exp \left[ \int d\tilde k~ |\tilde J(k)|^2 \left\{ f (z_0(k) - 1)
\right. \right. \\ 
  &   & \left. ~~~~~~\vphantom{\int} \left. + (1-f)(z_{ch}(k) - 1)
\right\} \right]
~. 
\end{eqnarray*}

Thus far we have averaged over chiral orientations of the classical
source, but not the overall shape of the source $J$. Eventually this
problem must of course be faced. If the fluctuations about the mean,
classical $J$ are Gaussian, there exists a well developed formalism for
dealing with them \cite{apw}. Indeed the DCC average we have performed
is in fact also essentially carried out in Ref. \cite{apw} by Andreev,
Pl\"umer, and Weiner.

In the case of DCC, the fluctuations will arguably go beyond the
Gaussian approximation even when the DCC fluctuations are generated by
a Gaussian distribution of initial condition parameters. This is a big
problem, beyond the scope of this paper. We will comment on it again in
the concluding section.

\section{Simple nonlinear sigma-model DCCs} 

\subsection{Setting up 
the problem in terms of an auxiliary field}

The picture of DCC evolution given in the sections 1 and 3 can be
implemented (although, at least at early times, one does not obtain an
accurate quantitative description of the physical system) within the
nonlinear sigma-model in the framework of the set of classical
solutions\cite{anselm,blaizot,kota}, which have the form
\begin{eqnarray}
\Phi = f_\pi \, V_f^\dagger e^{ i \theta \tau_3} V_f
~,\label{urotaz}
\end{eqnarray}
where $V_f$ is a constant but otherwise arbitrary unitary
matrix\cite{anselm}, which orients $\Phi$ along a direction in isospin
corresponding to neutral fraction $f$, and $\theta$ is such that
\begin{eqnarray}
        \Box \theta = 0 
~.\label{eqsteta} 
\end{eqnarray}
(We are for the moment specializing our analysis to the chiral 
limit $m_\pi \! = \! 0$.)

For $V_f \! = \! 1$ (which corresponds to $f \! = \! 1$)
these solutions describe chiral fields 
with $\pi_1 \! = \! \pi_2 \! = \! 0$ and
\begin{eqnarray}
& \pi^0 \!\! =  f_\pi \sin \theta \label{tetatopi} &\\
& \sigma \!\! = \sqrt{f_\pi^2 - (\pi^0)^2} = f_\pi \cos \theta &
~.\label{tetatosigma} 
\end{eqnarray}

Our analysis will exploit the simplicity of Eq.~(\ref{eqsteta}), and
will be based on the relation between source and particle flux
discussed in the previous section. One aspect that perhaps deserves
clarification is the relation between the pion flux and the solutions
of the evolution equation for the field $\theta$ in presence of a
$\theta$-source
\begin{eqnarray}
        \Box \theta(x) & = & J_\theta(x)
~. \label{evolteta}
\end{eqnarray}
For this, it is important to observe that when the source term
$J_\theta$ is well localized in space and time, as it must be for the
consistency of our approach, the amplitude of the field $\theta$ must
(because of energy conservation) become small everywhere at
sufficiently late times. Therefore the relation between the late-time
asymptotic fields $\pi^0_{out}$ and $\theta_{out}$, which according to
(\ref{tetatopi}) is given by\footnote{Note that the asymptotic behavior
of the fields and sources plays a rather central role in our analysis.
In particular, a class of solutions more general than (\ref{urotaz}),
given by $\Phi \! = \! f_\pi \, W^\dagger e^{ i \theta \tau_3} U$ with
$U \! \ne \! W$, could be considered, but would require everlasting
sources. Such idealized sources have been considered in the analysis of
Ref.\cite{lampert} (where the source never turns off as a result of
boost invariance) and Ref.\cite{blaizot} (where the source never turns
off as a result of the infinite size of the ``pancake'') and can be
useful in deriving some intuition about DCCs, but do not reproduce the
experimental conditions of DCC searches.} 
\begin{eqnarray}
\pi^0_{out}(x)  = f_\pi \, \sin \theta_{out}(x) ~, 
\label{exactrelout}
\end{eqnarray} 
can be well approximated by 
\begin{eqnarray}
\pi^0_{out}(x)  \approx f_\pi \, \theta_{out}(x) ~.
\label{approxrelout} 
\end{eqnarray}

Based on the analysis reported in the previous section, it is then easy
to see that the corresponding pion production is described by the
generating functional
\begin{eqnarray}
G[z_\pi] & = & \exp \left[ \int d\tilde k~ 
f_\pi^2 |\tilde J_\theta(k)|^2 (z_\pi(k) - 1) \right]
~, \label{generpizero}
\end{eqnarray}
and in particular, the inclusive spectrum of pions is given by
\begin{eqnarray}
2\omega_k {dN_{\pi^0} \over d^3 \vec k} & \approx & {f_\pi^2 \over
(2\pi)^3} \, |\tilde J_\theta(\vec k)|^2
~. \label{fluxtetatopion}
\end{eqnarray}

For $V_f \! \ne \! 1$, a simple generalization of this argument holds,
and the (charged and neutral) pion production is ultimately described
within the generating functional formalism discussed in the preceding
section.

\subsection{Pion flux for a class of sources}

We are finally ready to define in more precise terms a simple model
which embodies the essence of the baked-alaska scenario.
We start by introducing a source of the general form\footnote{The 
normalization factor has been chosen so that (as explicitly 
shown later on) for small enough $t$ the
DCC is pure $\pi = f_\pi$ inside the light-cone.}
\begin{eqnarray}
        J_\theta(x) & = & 4\pi^2 f(t) D_{ret}(x)
~, \label{genesource}
\end{eqnarray}
where 
\begin{eqnarray}
D_{ret}(x) & = & {1 \over 4\pi r} \Theta(t) \delta(r-t)
\end{eqnarray}
is the retarded Green function for a massless scalar field, and
$f(t)$ is a function of $t=x^0$ such that
\begin{eqnarray}
f(t) \rightarrow 0 & t \rightarrow \infty \nonumber\\
f(t) \rightarrow 1 & t \rightarrow 0
~, \label{genef}
\end{eqnarray}
{\it i.e.} the source is {\it on} at early times but eventually
switches {\it off}.
We are also using standard notations for $r \equiv \! |\vec x|$
and for the step function $\Theta$.

The physically interesting
quantity, as discussed in section \ref{classapp}, 
is $\tilde J_\theta (\vec k)$,
and the Fourier transform of the current density (\ref{genesource})
is given by
\begin{eqnarray}
\tilde J_\theta(k) & = & \int d^4x~e^{ikx} J_\theta(x) \nonumber\\
& = & \int dt~e^{ik^0t}~{4\pi \over \kappa}
\int^\infty_0 dr~r\sin \kappa r \left[ 
f(t) {\pi \over r} \Theta(t) \delta(r-t) \right] \nonumber\\
& = & {4\pi^2 \over \kappa} \int^\infty_0 dt~f(t) e^{ik^0t}
\int^\infty_0 dr~\delta(r-t) \sin \kappa r 
\label{genejresexact} \\
& = & {4\pi^2 \over \kappa} \int^\infty_0 dt~f(t) e^{ik^0t}\sin \kappa t 
~, \nonumber
\end{eqnarray}
where we used the notation $\kappa \! \equiv \! |\vec{k}|$.
It then immediately follows that
\begin{eqnarray}
\tilde J_\theta (\vec k) & = & {4\pi^2 \over \kappa} 
\int^\infty_0 dt~f(t) e^{i \sqrt{{\kappa^2 + m^2}} t} \sin \kappa t 
~. \label{genejresexactmasshell}
\end{eqnarray}
In particular, $\tilde J_\theta (\vec k)$
can be approximated at large $\kappa$ as
\begin{eqnarray}
\tilde J_\theta (\vec k) \approx {2\pi^2i \over \kappa} 
\int^\infty_0 dt~f(t) e^{i {m^2 \over 2 \kappa} t}
\sim {2\pi^2i T \over \kappa} 
~, \label{genejresapproxmasshell}
\end{eqnarray}
where $T \! \equiv \! \int dt~f(t)$. We see that the large $\kappa$
behavior is not very sensitive to the exact form of the function
$f(t)$. However, this does assume that the sources are localized
exactly on the light-cone. If we introduce a representation of the
delta function with a finite width, the resulting field distribution is
smoother and, as one might expect, the high frequency tail falls off
much more rapidly.

\subsection{Pion flux and field evolution for a specific source}

It is interesting to consider the idealized case $f(t) = \Theta(T-t)$,
where the source term switches off suddenly at some time $t \! = \! T$,
to be associated with the decoupling time. In this case the source
takes the form
\begin{eqnarray}
J_\theta(x) & = &  \Theta(T-t) {\pi \over r} \Theta(t) \delta(r-t) 
~, \label{simplesource}
\end{eqnarray}
and Eq.~(\ref{genejresexact}) 
describing $\tilde J$ off the mass shell
reduces to
\begin{eqnarray}
        \tilde J_\theta(k) & = & {4\pi^2 \over \kappa} \int^T_0
dt~e^{ik^0t}\sin \kappa t \nonumber \\
                    & = & -i~{2\pi^2 \over \kappa} \left\{
{e^{iT(k^0+\kappa)}-1 \over i(k^0+\kappa)} - {e^{iT(k^0-\kappa)}-1 \over
i(k^0-\kappa)}\right\} 
~. \label{stepfnsol}
\end{eqnarray}
The value of $\tilde J$ on the $k^0 \! > \! 0$ mass shell
is easily obtained by substituting
$\omega_k \! = \! \sqrt{\kappa^2+m^2}$
for $k^0$
and, in particular, in the massless case 
(and the large-$\kappa$ limit of the massive case)
one finds
\begin{eqnarray}
\tilde J_\theta(\vec k) & = & -i~{2\pi^2 \over \kappa} 
\left\{ {e^{i2T \kappa} - 1 \over 2 i \kappa} - T\right\} \nonumber\\
& = & -i~{\pi^2 \over \kappa^2} \left\{ -2T \kappa +
i(1-e^{i2T \kappa}) \right\} 
~. \label{masslcase}
\end{eqnarray}

For the simple choice of source considered in this subsection
it is also possible to derive explicitly
the corresponding $\theta$-field.
We start by observing 
that 
\begin{eqnarray}
\theta(x) & = & \int d^4{x^\prime}~D_{ret}(x-{x^\prime})
J_\theta({x^\prime}) \nonumber\\ 
& = & \int d^4{x^\prime}\left\{ {\Theta(t-{t^\prime}) \over
4\pi|\vec x - \vec {x^\prime}|} \delta(|\vec x - \vec
{x^\prime}|-(t-{t^\prime})) \right\} \cdot
\Theta(T-{t^\prime})\left\{ {\pi\Theta({t^\prime} ) 
\over {r^\prime}} \delta({r^\prime} - {t^\prime}) \right\} \nonumber \\
                  & = & {1 \over 4} \int
d^4y~\Theta(T-{t^\prime})\Theta({t^\prime})\Theta(t-{t^\prime})\left({1
\over {r^\prime}|\vec x - \vec y|}\right) \delta(|\vec x - \vec
{x^\prime}|-(t-{t^\prime}))~\delta({r^\prime} - {t^\prime}) \nonumber \\
                  & = & {1 \over 4} \int
d^4y~\Theta(T-{t^\prime})\Theta({t^\prime})\Theta(t-{t^\prime})\left({1
\over {r^\prime}\rho} \right) \delta(\rho
-(t-{t^\prime}))~\delta({r^\prime} - {t^\prime}) 
~, \label{eqteta1}
\end{eqnarray}
where in the last line we have introduced the variable $\rho=|\vec x
-\vec x^\prime|$.
We now go to spherical polar coordinates, choosing the $z$-axis to lie
along $\vec x$ so that $\rho=\sqrt{r^2 + r^{\prime 2} -
2rr^\prime\mu}$, and find that
\begin{eqnarray}
\theta(x) & = & \!\!\!{\pi \over 2} \int^T_0 \! d{t^\prime} \, \Theta(t-{t^\prime})
\int^\infty_0 \! dr^\prime \delta({r^\prime} - {t^\prime}) \int^1_{-1}
\! d\mu \, {r^\prime \over \rho} \delta(\rho
-(t-{t^\prime})) \nonumber\\ 
        \!\!\! & = & \!\!\!{\pi \over 2r} \int^T_0 \! d{t^\prime} \,
\Theta(t-{t^\prime}) \int^\infty_0 \! dr^\prime \delta({r^\prime} -
{t^\prime}) \int_{|r-r^\prime|}^{r+r^\prime}\! d\rho \, \delta(\rho
-(t-{t^\prime})) \nonumber\\ 
        \!\!\! & = & \!\!\!{\pi \over 2r} \int^T_0 \! d{t^\prime} \,
\Theta(t-{t^\prime}) \int^\infty_0 \! dr^\prime \delta({r^\prime} -
{t^\prime}) \Theta((t-{t^\prime})-|r-r^\prime|) \Theta(r+r^\prime -
(t-{t^\prime})) \nonumber\\ 
        \!\!\! & = & \!\!\!{\pi \over 2r} \int^T_0 \! d{t^\prime} \,
\Theta(t-{t^\prime}) \Theta((t-{t^\prime})-|r-t^\prime|)
\Theta(2t^\prime - t + r) \nonumber\\ 
        \!\!\! & = & \!\!\!  {\pi \over 2r} \int^T_0 d{t^\prime}~\Theta(t-{t^\prime})
\left\{\Theta(t-r)\Theta(2{t^\prime}-t+r) 
- \Theta(2{t^\prime}-t-r)\right\}
~. \label{eqteta3}
\end{eqnarray}
Here the last line (\ref{eqteta3}) follows after some straightforward
manipulation of the step functions in the integrand.
In order to render explicit the structure of the result (\ref{eqteta3})
it is convenient to examine separately the cases $t \! < \! 0$,
$0 \! < \! t \! < \! T$, and $T \! < \! t$; this also allows us
to see how the various stages of the evolution of DCC described in
the sections 1 and 2 are realized within this solution.
From Eq.~(\ref{eqteta3}) it follows that

\begin{figure}[htbp]
\epsfxsize=6.5in
\epsfysize=7.5in
\centerline{\epsffile{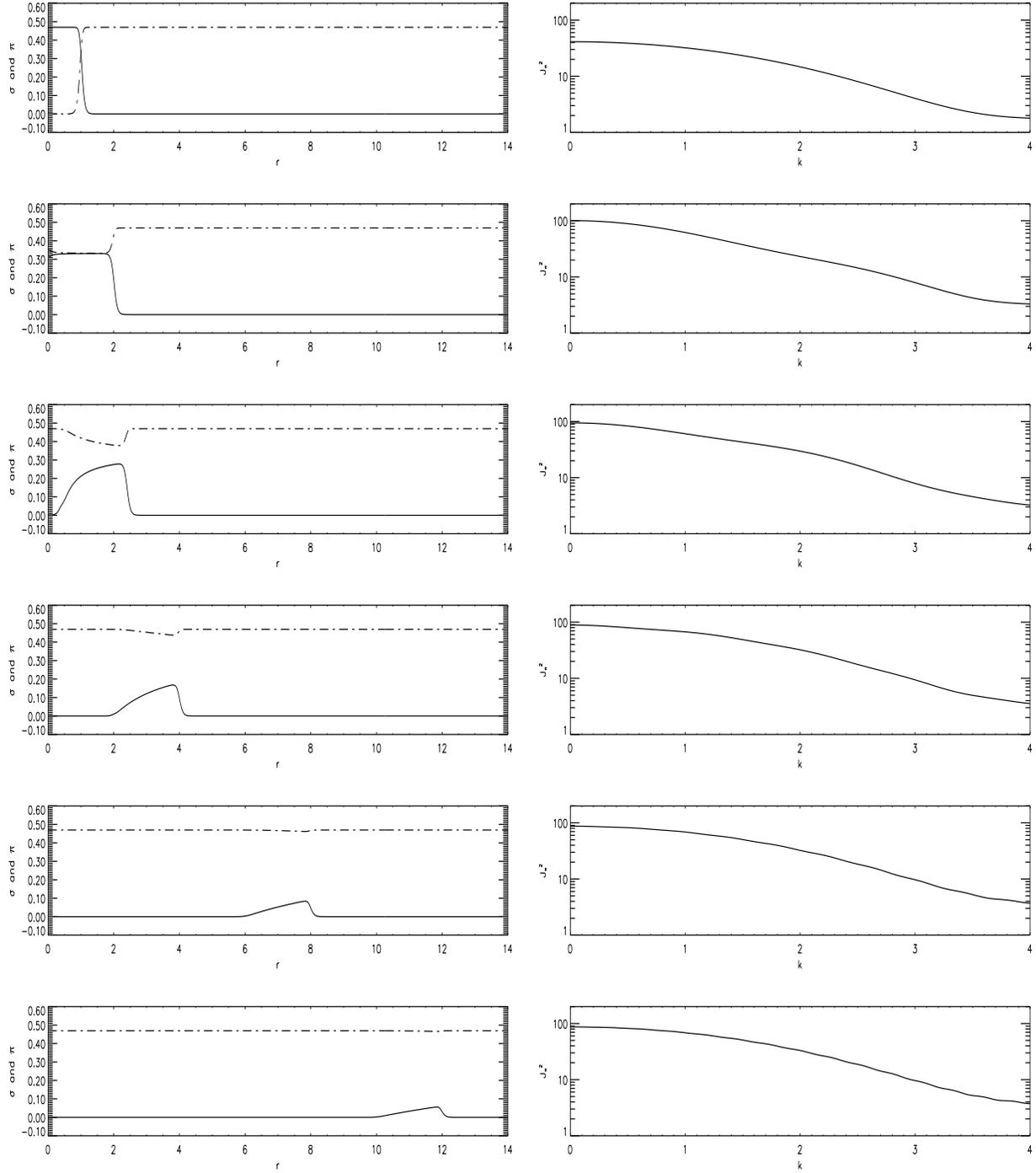}}
\caption[*]{Evolution of the
$\pi$ (continuous line) and $\sigma$ (broken line) fields according
to the classical nonlinear sigma-model, starting from pure 
DCC $\pi = f_\pi$ inside the light-cone at $t = 1 fm$.
The fields are shown in the snapshots on the left,
whereas the snapshots on the right
show the modulus squared of the pion source function (the $\tilde J$
of Eq.~(\ref{Jfromfield})). The horizontal scales for the left-side
and right-side plots are in fermi and $($fermi$)^{-1}$ respectively.}
\label{fig1}
\end{figure}

\begin{itemize}

\item
$\theta \! = \! 0$ 
when $t \! < \! 0$, as expected
since $D_{ret}(x)$ and $J_\theta(x)$ vanish for $t \! < \! 0$.

\item
For $0 \! < \! t \! < \! T$ 
\begin{eqnarray}
\theta(t,r) & = & {\pi \over 2} \Theta(t-r)
~, \label{earlydcc}
\end{eqnarray}
{\it i.e.} a uniform region of DCC is present inside the light-cone
during the interval when the source is {\it on}.

\item
Finally for times later than $T$ 
\begin{eqnarray}
\theta(t,r) & = & {\pi \over 4}\left({2T-t+r \over
r}\right)\Theta(2T-t+r)\Theta(t-r) \nonumber\\
                    &    & -{\pi \over 4}\left({2T-t-r \over
r}\right)\Theta(2T-t-r)
~, \label{eq8}
\end{eqnarray}
which is of the form $[f(t-r) + g(t+r)]/r$ and therefore satisfies the
free wave equation, as expected since in this region the current
density is {\it off}. The Eq.~(\ref{eq8}) also explicitly shows that at
$t \! = \! T$, when the source is {\it turned off}, the ordinary vacuum
(the $\sigma$-vacuum) starts breaking into the interior of the light
cone. At $t \! = \! 2T$ the ordinary vacuum reaches the small-$r$
region, and for times later than $2T$ the Eq.~(\ref{eq8}) can be
interpreted as describing the outward propagation, in ordinary vacuum,
of a localized DCC-wave. This sequence of events can be seen in the
$\pi$ and $\sigma$ fields plotted in Fig.~\ref{fig1}.

\end{itemize}

Having obtained the nonlinear sigma-model solution corresponding to the
source (\ref{simplesource}), we can use it for a {\it consistency
check} for the relation between source and field discussed in section
\ref{classapp}, {\it i.e.} we can check that using
Eq.~(\ref{Jfromfield}) one can indeed obtain the Fourier transform of
the source (\ref{simplesource}) from the late-time $\theta$-field,
which according to (\ref{eq8}) is given (fot $t>2T$) by\footnote{Notice
that the late-time $\theta$-field satisfies $|\theta_{out}(t,r)| \leq
\pi T / (2 t)$, and therefore the approximation $\pi_{out} \equiv f_\pi
\sin \theta_{out} \approx f_\pi \theta_{out}$ holds asymptotically, as
argued at the beginning of this section based on general
energy-conservation arguments.} 
\begin{eqnarray} 
\theta_{out}(t,r) & =
&  {\pi \over 4}\left({2T-t+r \over r}\right)\Theta(2T-t+r)\Theta(t-r)
~, \label{eq8late} 
\end{eqnarray} 
We start by evaluating the three
dimensional Fourier transform of $\theta_{out}$ 
\begin{eqnarray}
{\tilde \theta}^{(3)}_{out}(t,\kappa) & = & {4\pi \over \kappa}
\int^\infty_0 dr~r \sin \kappa r~{\pi \over 4}\left({2T-t+r \over
r}\right)\Theta(2T-t+r)\Theta(t-r) \nonumber\\ 
& = &  {\pi^2 \over
\kappa} \int^\infty_0 dr~(2T-t+r) \sin \kappa
r~\Theta(2T-t+r)\Theta(t-r) \label{foureq8late} \\ 
& = &  {\pi^2 \over
\kappa} \int^t_{t-2T} dr~(2T-t+r) \sin \kappa r \nonumber\\
        & = &   {\pi^2 \over \kappa^2} \left[ -2T \cos \kappa t 
+ {\sin \kappa t \over
\kappa} + {\sin \kappa (2T-t) \over \kappa} \right]
~. \nonumber
\end{eqnarray}   
The three dimensional Fourier transform of $\dot\theta$ is then given by
\begin{eqnarray}
\tilde{\dot\theta}\vphantom{\theta}^{(3)}(t,\kappa) & 
= & {d \over dt} {\tilde \theta}^{(3)}(t,\kappa)
\nonumber\\
& = &  {\pi^2 \over \kappa} \left[ 2T \sin \kappa t + {1 \over \kappa}(\cos
\kappa t - \cos \kappa (2T-t)) \right]
~. \label{dotfoureq8late} 
\end{eqnarray}
Finally, following Eq.~(\ref{Jfromfield}) we get 
(also taking into account that this calculation is in the massless
limit, and therefore $\omega_k = \kappa$)
\begin{eqnarray}
\tilde J_\theta(\vec k) & = & -i~e^{i \kappa t}\left[ \kappa \tilde
\theta^{(3)}_{out}(t,\kappa) 
+ i\dot{\tilde \theta}\vphantom{\theta}^{(3)}_{out}(t, \kappa)\right]
\nonumber\\
        & = & -i~~e^{i \kappa t}~{\pi^2 \over \kappa}\left[ -2T 
e^{-i \kappa t} + {i \over
\kappa}e^{-i \kappa t} - {i \over \kappa}e^{i \kappa (2T-t)}\right]
\label{lasteqsec4} \\
        & = & -i~{\pi^2 \over \kappa^2} \left[ -2T \kappa + i \left( 1 
- e^{i2T \kappa}
\right) \right] 
~, \nonumber
\end{eqnarray}
in complete agreement with Eq.~(\ref{masslcase}).

In Fig.~1 we show a few snapshots of the evolution
of the fields for the solution (\ref{eqteta3})
and the corresponding source obtained using Eq.~(\ref{Jfromfield}),
which agrees very well with the expression (\ref{masslcase}).

\subsection{Correlation between generic and DCC pion production}

The question of correlations between generic and DCC pion production is
very important for experimental DCC searches. It is quite reasonable
that such correlations should exist. In our baked-alaska scenario, if
the source strength on the light cone is large, {\it i.e.}~lasts for a
long time $T$, then the amount of DCC which is produced will, as we
have seen, also be large. To be more quantitative is not easy. Here we
attempt to make the connection by a simple argument based on energy
densities. In the absence of the source, then there would be a
surface-tension, {\it i.e.}~an energy per unit area, associated with
the boundary region between the sphere containing the DCC and the
normal vacuum on the outside. It is contributed by the kinetic-energy
term of the DCC hamiltonian, since gradients of the pion field exist in
the interface region. It is reasonable to assume that when the energy
per unit area of the generic partons or hadrons in the source region
exceeds this surface energy, the inner DCC will be decoupled from the
outer vacuum, and conversely when this energy is less than the DCC
surface-tension, the source term will be inoperative.\footnotemark~This
will allow a connection between the amount of generic production and
the amount of DCC produced.

\footnotetext{It may also be argued \cite{privBirse} that the source of
the pion field should be a 
scalar density built from the constituent quarks composing the generic
material on the light-cone. This quantity is not the energy-momentum
tensor, so that this is not obviously the same criterion as we are using.
On the other hand, when the fields are rapidly varying (as they are in the
source region), it is not clear what the correct choice of source of pion
degrees of freedom is, and a simple argument based on energetics seems
not unreasonable to try.}

In the spherically symmetric case, by assumption the DCC surface
energy is contained 
in a shell of thickness $\Delta$, with $\Delta$ a typical hadronic
scale, say $0.2fm < \Delta < 0.6fm$. We can therefore write
\begin{eqnarray}
{E_{DCC} \over A} \sim {1 \over 4 \pi T^2}
\int^{T}_{T-\Delta} d r \, 4\pi r^2 {1 \over 2} f_\pi^2 (\nabla \theta)^2
\sim {f_\pi^2 \over 2 \Delta}
~, \label{edordi}
\end{eqnarray}
where the last approximation on the right-hand side follows from
approximating the profile at decoupling of the $\theta$-field near $r =
T$ with a linear interpolation between $\theta(r=T-\Delta,t=T)=1$ and
$\theta(r=T,t=T)=0$.

For the energy per unit area of the hot shell made of collision debris,
one easily finds
\begin{eqnarray}
{E_{shell} \over A} = 
{dE_{shell} \over dA} \sim
\left[{ <p_T> \, {dN  \over d\Omega} d\Omega \over T^2
d\Omega}\right]_{generic}
= \left[{ <p_T> \over T^2} {dN \over d\Omega}\right]_{generic} 
~. \label{eddcc}
\end{eqnarray}

If indeed at $t=T$ the DCC surface energy density equals the energy per
unit area of the hot shell of collision debris, one finds from
(\ref{edordi}) and  (\ref{eddcc}) that
\begin{eqnarray}
\left[<p_T> {dN \over d\Omega}\right]_{generic} = 
T^2 {f_\pi^2 \over 2 \Delta}
~. \label{corr1}
\end{eqnarray}

The aforementioned correlation between generic and DCC pion production
is then seen upon combining (\ref{corr1}) with (\ref{fluxtetatopion})
and (\ref{genejresapproxmasshell}). Specifically, one finds (in the
chiral limit)
\begin{eqnarray}
\left[p {dN \over d\Omega dp}\right]_{DCC} 
\sim {2 \pi \over \Delta} \left[<p_T> {dN  \over d\Omega}\right]_{generic} 
~. \label{eeee}
\end{eqnarray}

We therefore see that it is possible for DCC production and generic
production to be comparable in terms of the number density of produced
particles.

\section{Beyond the classical nonlinear sigma-model}

Up to this point we have based our description of DCCs on linear,
semiclassical, coherent-state solutions of a simple nonlinear
sigma-model. However, there has been a considerable body of work on DCC
production which goes substantially further. Not only are the nonlinear
equations of the linear sigma-model (or even more complicated
models~\cite{quarks}) considered, but also the effects of quantum
fluctuations are taken into account within the
mean-field~\cite{BdVH,gacCSM} (or large-$N$~\cite{BdVH,CKMP})
approximation. This level of calculation has become the de facto
``state-of-the-art''.  However, in most cases the space-time geometry
is greatly simplified, or else the approach is aggressively numerical.

The closest calculation at this level to what we have presented here
has been performed by Lampert, Dawson and Cooper (LDC) \cite{lampert}.
They consider a boost-invariant spherical expansion, such that the
fields depend only upon the proper time which has elapsed since the
expansion began. This is not very realistic, because the inclusive
particle distribution which emerges must be the same in all reference
frames, and therefore requires an infinite mean energy per particle,
and an infinite formation-time for the final-state distribution.

While the LDC solutions are, as they stand, not very realistic, it is
not too hard to adapt them to the baked-alaska scenario which we have
described. We shall sketch in subsequent subsections how this works.
The main point is that if we assume that the dynamics is as described
in LDC for times $t$ less than $T$, after which time the sources on and
near the light cone are turned off, then by causality the LDC solution
will still be exact within the double-cone region, {\it i.e.}  between
the forward light-cone (vertex at $t=0$) and an inverted light cone
with vertex at $t=2T$. If $T$ is large enough, one could hope that the
fields inside the double-cone region and far from the light-cone be
asymptotic, so that they could be matched onto the nonlinear
sigma-model fields we use, and an estimate of the low-momentum portion
of the particle spectrum could then be obtained using
Eq.~(\ref{Jfromfield}). The quality of this method can be tested by
varying $T$ and determining which portion of the inclusive spectrum is
insensitive to $T$.

For a more rigorous analysis of the spectrum, the source should be
turned off at the decoupling time $T$, but the fields should be evolved
according to the linear sigma-model up to times late enough for the
evolution to be effectively free. At sufficiently late times one can
then reliably extract the information on the inclusive spectrum using
the procedure described in Sec. 4.

It is also instructive to consider the simplification achieved by
neglecting the mean-field quantum corrections, in which case the LDC
calculation is reduced to the solution of coupled ordinary differential
equations describing the evolution of the classical fields in proper
time. We have made such calculations for initial conditions chosen by
LDC, and find remarkably close agreement of the time dependence of the
pi and sigma fields with what is obtained from the full mean-field
quantum calculation. This is encouragement that, when one goes on to
consider the linear sigma-model in more difficult, less symmetric
geometries, a classical calculation may well suffice to provide at
least a semiquantitative picture of the dynamics which the mean-field
calculation would provide. After all, only a semiquantitative
description need be obtained from the linear sigma-model, because it is
just a rough approximation to the complete low-energy effective chiral
Lagrangian of {\it real} QCD.

In the following subsections we sketch more details of this line of
argument. In subsection 5.1 we discuss the connection between the pion
flux as calculated from the nonlinear sigma-model with what one would
obtain from a classical solution of the linear sigma-model. In
subsection 5.2 we describe in more detail the LDC analysis, especially
in the classical approximation, and we numerically compare the
mean-field and classical LDC-type solutions. In subsection 5.3 we match
the classical versions of the LDC solutions at time $T$ onto the free
asymptotic fields of the nonlinear sigma-model, thereby defining an
effective source function $J$, from which the pion distributions are
calculated. Finally, in subsection 5.4 we match the LDC solutions at
time $T$ onto fields of the linear sigma-model, evolve according to the
linear sigma-model up to some time $T^\prime$ say (late enough for the
evolution to be effectively that of a free field) when the calculated
effective source function will represent the actual pion flux.

\subsection{Deriving the pion flux in more general frameworks}

We start by applying our formalism to the derivation of the pion flux
for the full linear sigma-model or related models, rather than for the
nonlinear sigma-model considered in the previous sections. Provided one
is dealing with a well-defined scattering problem, with the sources
localized in space and time, the evolution of the $\pi$ and $\sigma$
degrees of freedom will eventually reach an asymptotic regime governed
by free field behavior. An estimate of the associated inclusive pion
spectrum can indeed be obtained by applying the formulas discussed in
the previous sections.

While technically this procedure is rather straightforward, it is
important to realize that the associated ``sources'' are somewhat
different from the ones we have been discussing. In this more general
case one is actually dealing with ``effective sources'', useful as
computational tools in the analysis of pion production, but not to be
interpreted as physical external sources in the problem. From the point
of view of the original model, say the linear sigma-model, these
effective sources are given by the sum of a physical external source
and a term from the self-interactions of the fields.

These ideas are of rather general applicability; for example,
in the investigation of an interacting system
described by the Lagrangian density
\begin{eqnarray}
{\cal L} = {1 \over 2} (\partial_\mu \Phi) (\partial^\mu \Phi)
- {m^2 \over 2} \Phi^2 - V(\Phi)
~ \label{lagen}
\end{eqnarray}
one is naturally led to the study of the evolution equation
\begin{eqnarray}
(\Box + m^2) \Phi = -V'(\Phi) + J
~, \label{boxie}
\end{eqnarray}
where $J$ is a physical ``external'' ($\Phi$-independent) source
and $J_{\it eff} \equiv -V'(\Phi) + J$
is an effective source.

\begin{figure}[htbp]
\epsfxsize=6.5in
\epsfysize=7.5in
\centerline{\epsffile{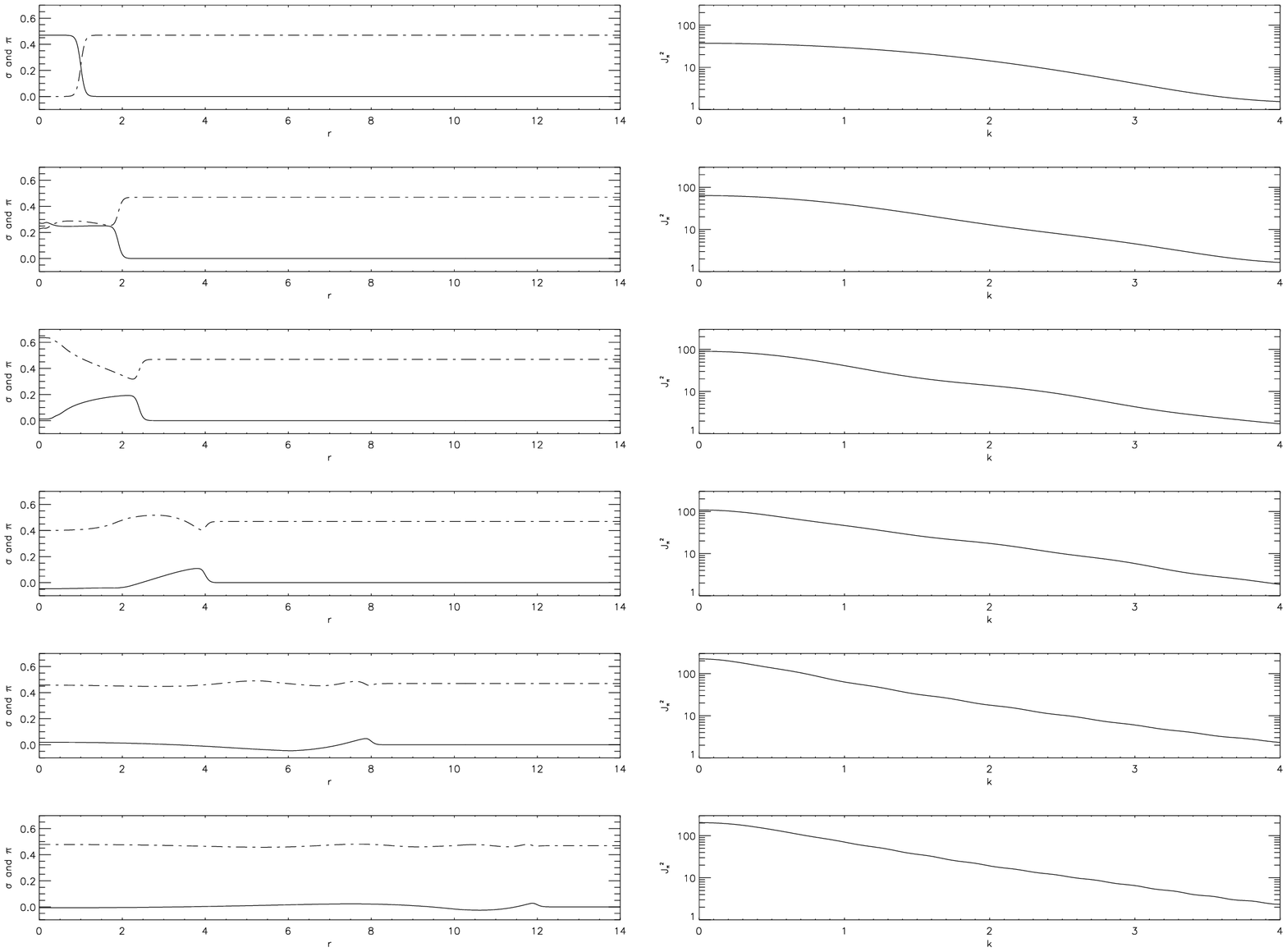}}
\caption{Evolution of the $\pi$ (continuous line) and $\sigma$ (broken
line) fields,
and the pion source function according
to the linear sigma-model, starting from pure 
DCC $\pi = f_\pi$ inside the light-cone at $t = 1 fm$.}
\label{fig2}
\end{figure}

The simulation reported in Fig.~\ref{fig2} corresponds to the linear
sigma-model classical evolution from an initial configuration given by
pure DCC, $\pi = f_\pi$, inside the light-cone and true vacuum outside
({\it i.e.} a snapshot of the solution (\ref{earlydcc}) at some chosen
time) and vanishing initial field velocities (except on the light
cone). We simulate the classical field equations, as obtained from the
Lagrangian density (\ref{lagrlin}), for a spherically symmetric field
configuration. Rather than simulating the $\Phi$ field directly, our
program evolves $r \, \Phi(t,r)$ which simplifies the form of the
d'Alambertian. The boundary conditions at the origin are set up to
ensure that $\Phi(t,r)$ is an even function of $r$ at all times. The
relevant classical field configuration is then evolved in time, using a
simple staggered leapfrog algorithm (see for example \cite{NRinC}). The
Fourier transforms involved in finding the source current $\tilde
J(\vec k)$ as defined in expression (\ref{Jfromfield}) are done using a
straightforward (extended trapezoid) method.\footnote{Note that by
setting $\lambda=H=0$ in (\ref{lagrlin}) we obtain the free wave
equation. This enables us to use an almost identical program to evolve
the $\theta$-field in the nonlinear sigma model to obtain the graphs in
Fig.~\ref{fig1}. The major change in the code between the two cases is
in fact the use of the identity $\pi=f_\pi\sin\theta$ in the $\tilde
J$-extracting routine.} Fig.~\ref{fig2} shows a selection of output
snapshots chosen to illustrate the main features of the evolution. The
pictures on the left describe the evolution of the pion and sigma
fields, while the pictures on the right describe the corresponding
``evolution'' of the Fourier-space effective source function. The
emergence of a stationary Fourier-space effective source (which encodes
the information on particle production) reflects the fact that at late
times the evolution of the $\pi$ and $\sigma$ degrees of freedom
reaches an asymptotic regime ruled by the nonlinear sigma-model.
However, it should be noted that the low momentum part of the spectrum
is only complete after a time scale between $4$ and $8$ fermi, and is
significantly larger (a factor of $2$) than what was obtained for the
nonlinear sigma-model for identical initial conditions.

\subsection{Classical version of LDC approach
and reliability of coherent-state descriptions}

Our description of DCCs uses the classical equations of motion to
obtain an ``out'' field from a given ``in'' field. This ``out'' field
is then mapped into a corresponding coherent state from which particle
production (a quantum effect) can be derived. Some elegant recent
studies~\cite{BdVH,CKMP,lampert} have been based on more general
formalisms for the description of quantum effects and have taken into
account (some of) the non-perturbative quantum effects contributing to
the structure of the full propagator. One is then confronted with the
solution of a genuinely non-classical evolution problem, in which ``gap
equations'' describing the full propagators are combined with
(modified) evolution equations for the fields. Seen as solutions of a
variational problem, these equations result from finding an extremum of
the (quantum) effective action for composites\cite{cjt}, just like the
classical evolution equations are obtained from finding an extremum of
the classical action for (local/non-composite) fields.

In Ref.\cite{lampert}, LDC investigated the chiral phase transition by
modeling the relevant hadron dynamics with a linear sigma-model, and
adopting evolution equations that take into account part of the
non-perturbative quantum effects contributing to the structure of the
full propagator via the familiar large-$N$ formalism. They concentrated
on boost-invariant spherical expansions, such that the mean-field
expectation values depend only upon the proper time $\tau \! = \!
\sqrt{t^2 - r^2}$. We refer the interested reader to Ref.\cite{lampert}
for the complete description of the LDC approach. For the purposes of
the analysis presented in the remainder of this section, it is
sufficient for us to consider explicitly the evolution equations
corresponding to the classical version of the LDC equations, {\it i.e.}
obtained from the LDC equations by dropping all contributions coming
from the dressing of the propagator:
\begin{eqnarray}
\left[ {1 \over \tau^3} {\partial \over \partial \tau}
\left( \tau^3 {\partial \over \partial \tau} \right) 
+ \lambda ( \sigma^2 + {\bf \pi}^2 - f_\pi^2) \right] \sigma = f_\pi m_\pi^2
\label{sigmalam}\\
\left[ {1 \over \tau^3} {\partial \over \partial \tau}
\left( \tau^3 {\partial \over \partial \tau} \right) 
+ \lambda ( \sigma^2 + {\bf \pi}^2 - f_\pi^2) \right] \pi = 0
~. \label{pilam}
\end{eqnarray}

As a preliminary test of the reliability of our description of DCCs
based on the classical evolution equations, we have compared the
results for the mean-field evolution of the $\pi$ and $\sigma$ fields
obtained in Ref.\cite{lampert}, to the corresponding results from the
classical equations (\ref{sigmalam})-(\ref{pilam}). In Fig.~3 the
results of this comparative analysis are reported for the initial
conditions singled out in Ref.\cite{lampert} as the most
``DCC-favoring'' within the special family of initial conditions
considered there; specifically, we integrate the fields from the
initial conditions
\begin{eqnarray}
	\sigma  =  0~fm^{-1}~~ & & \dot \sigma  = -1~fm^{-2} 
\nonumber\\
	\pi  = 0.3~fm^{-1} & & \dot \pi  =  0~fm^{-2}
\label{initcondy}
\end{eqnarray}
at $\tau=1~fm$. Since the focus of this exercise is only on the field
evolution, rather than particle production, for simplicity we kept the
exact (spherically-symmetric and boost-invariant) source structure
adopted in Ref.\cite{lampert}. Fig.~3 suggest that even at a
quantitative level the description might be satisfactorily accurate.
This is especially so because one is in any case using rough models of
the relevant hadron dynamics ({\it i.e.}~it appears to be likely that
the inaccuracies introduced by using classical evolution equations
might be less important than the ones resulting from modeling the
relevant hadron dynamics with, say, the linear sigma-model).

\begin{figure}[htbp]
\epsfxsize=5.4in
\epsfysize=6.2in
\centerline{\epsffile{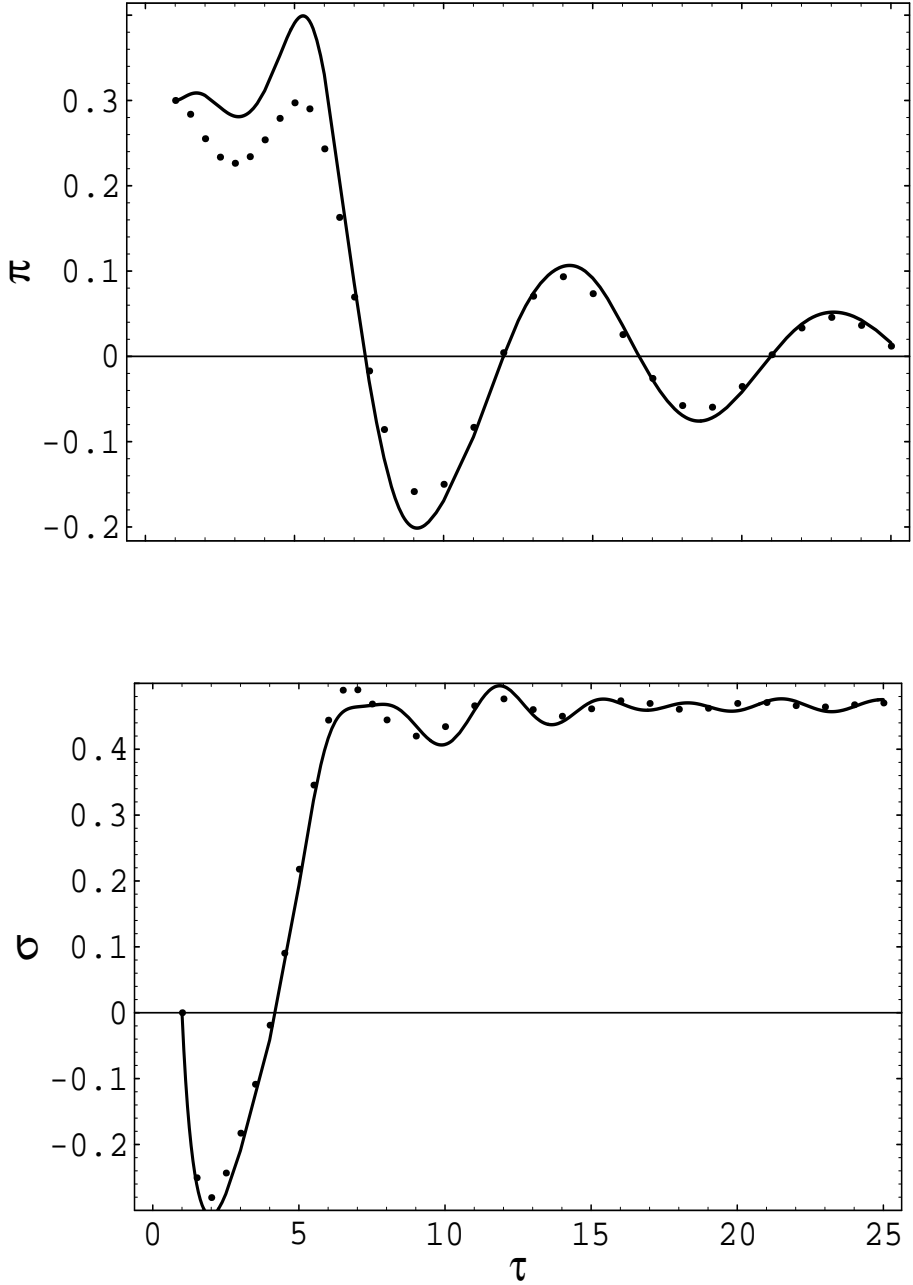}}
\caption{Evolution of the pion and sigma fields as functions 
of proper time in the LDC setup.
The continuous line corresponds to the purely classical analysis,
whereas the dotted line corresponds to the quantum analysis
reported in 
Ref.~[18]. 
At proper time $\tau \! = 1 fm$ 
the fields and their derivatives are fixed to
be $\pi = 0.3 fm^{-1}$, $\sigma \! = \! 0 fm^{-1}$, 
$\dot \pi \! = \! 0 fm^{-2}$, $\dot \sigma \! = \! - 1 fm^{-2}$.}
\label{fig3}
\end{figure}

\subsection{Low-momentum portion of the inclusive
spectrum in the LDC approach}

As explained at the beginning of this section, the spherical expansion
investigated in Ref.\cite{lampert}, which is fueled by everlasting
sources, and involves fields depending upon only the proper time, is
not very realistic. Still, as mentioned above one could attempt to
extract the low-momentum portion of the inclusive pion spectrum
associated with a baked-alaska-type modification of the LDC approach,
by mapping the LDC solutions at time $T$ onto free asymptotic fields of
the nonlinear sigma-model, thereby defining an effective source
function $J$, from which the pion distributions are calculated.
Ideally, one might find that for large enough $T$ the fields inside the
double cone region be almost everywhere asymptotic, and that only the
high-momentum tail of the particle spectrum could not be captured by
such an approach. In practice, however, we find that not even the
portion of the inclusive pion spectrum with very low momentum is well
determined at times as late as $50~fm$.

In Fig.~4 we report the results of one such analysis, in which we
integrate the fields from the initial conditions (\ref{initcondy}) at
$\tau=1~fm$, to some later time $T$ and use the boost symmetry to
reconstruct the fields everywhere on the $t=T$ surface inside the light
cone. $\tilde J(\vec k)$ is then extracted from this field
configuration using Eq.~(\ref{Jfromfield}) as before. The $\sigma$ and
$\pi$ field configurations and the effective source function described
for various choices of the above-mentioned time $T$ in Fig.~4 clearly
reflect the shortcomings of the approach discussed in this subsection.

\begin{figure}[htbp]
\epsfxsize=6.5in
\epsfysize=7.5in
\centerline{\epsffile{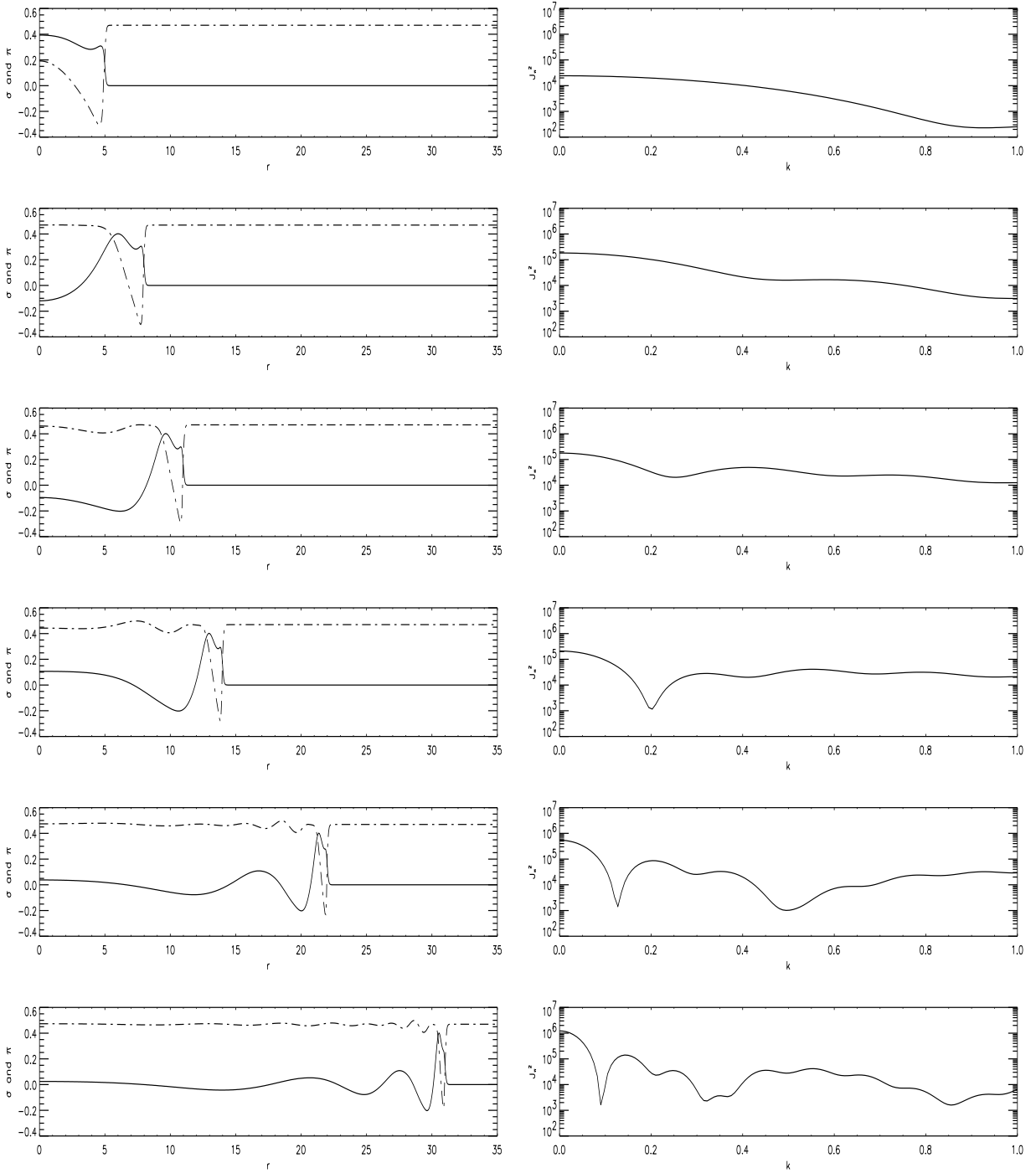}}
\caption{Evolution of the $\pi$ (continuous line) and $\sigma$ (broken
line) fields,
and the pion source function according
to the classical linear sigma-model, in the presence of a boost-invariant 
and spherically invariant source considered in 
Ref.~[18]. 
Again, the initial conditions 
are chosen according to the ``DCC-favoring'' scenario
considered in 
Ref.~[18].}
\label{fig4}
\end{figure}

\subsection{LDC approach with truncated sources}

The deficiencies of the method discussed in the previous section can be
easily remedied. Evidently, after the external source is turned off at
the decoupling time $T$, the fields should be evolved according to the
linear sigma-model up to times late enough for the evolution to be
effectively ruled by free field behavior. At such late times one can
reliably extract the information on the inclusive spectrum using the
procedure described in Sec.~4.

In Fig.~5 we report the results of such a simulation, again as
snapshots describing the evolution of the fields and the effective
source function corresponding to the initial configuration singled out
as ``DCC-favoring'' in Ref.\cite{lampert}. For illustrative purposes we
chose in this simulation a large decoupling time ($5 fm$). In this case
the effective source does reach a stationary regime; however, by
comparison with Fig.~2 we see that this asymptotic behavior only
emerges at rather late times ($\sim 20-30~fm$). The comparison of
Fig.~4 and 5 shows that the approach discussed in the preceding
subsection, in which the LDC sources were never turned off, can largely
overestimate ({\it e.g.} by more than a factor 10 if $T \le 5 fm$, as
assumed in Fig.~5) even the low-momentum portion of the spectrum.

\begin{figure}[htbp]
\epsfxsize=6.5in
\epsfysize=7.5in
\centerline{\epsffile{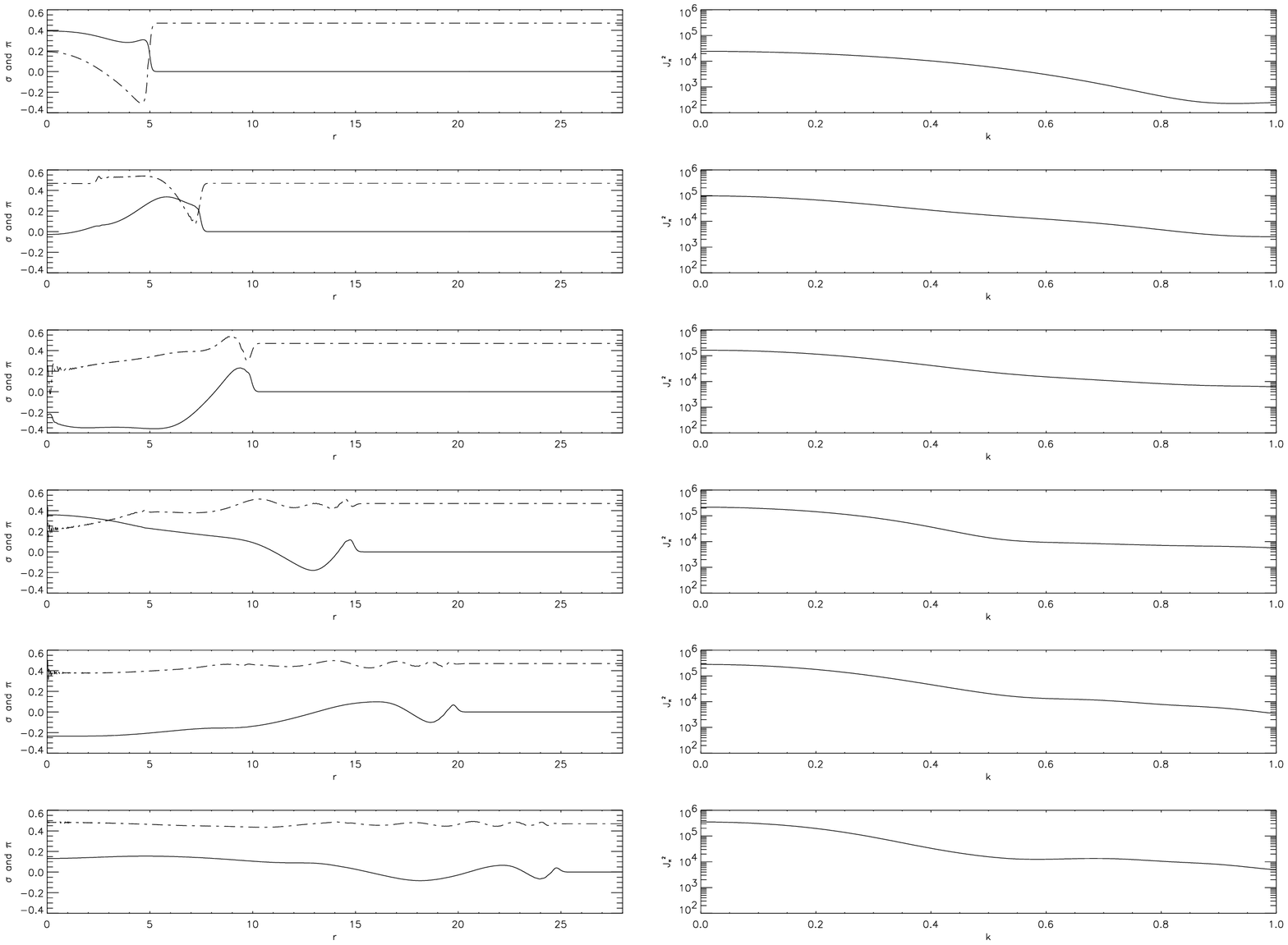}}
\caption{Evolution of the $\pi$ (continuous line) and $\sigma$ (broken
line) fields,
and the pion source function according
to the linear sigma-model, in the presence of a truncated version (as 
described in the text) of the
a boost-invariant 
and spherically invariant source considered in 
Ref.~[18]. 
The source is switched off at $t \! = \ 5 fm$.}
\label{fig5}
\end{figure}

\section{Conclusions}

In this paper we have investigated the various stages of the evolution
of disoriented chiral condensates via the ``baked-alaska'' mechanism.
Most of our analysis has been elaborated using classical equations of
motion based on either the linear or nonlinear sigma model. The
associated framework of coherent states was then used to make the
connection with the distribution functions for the particle production.
Important in this step is the identification of the source-function of
the produced particles, namely the right-hand side of the usual wave
equation (cf. Eq.~(\ref{boxie})). The square of the on-shell fourier
transform of this source function, as determined from the solutions of
the equations, then provides directly the inclusive distribution of
particles.

In general the source term consists of two parts. One is concentrated
near the light cone, and is a genuine external source, not a function
of the chiral fields, to be associated with the generic collision
debris of partons, constituent quarks, etc. The other part consists of
the nonlinear terms, built from the chiral fields themselves, which
appear in the classical wave equation. We found evidence, within the
nonlinear sigma-model approach, that the number of produced DCC pions
is likely correlated with the number of generic hadrons produced, with
this correlation local in (lego) phase-space. The number of DCC pions
could be comparable with the number of generic hadrons according to
this crude estimate, but the uncertainties are very large.

We also compared our very simple classical approach with a mean-field
calculation which includes one class of quantum corrections, and at
least in the case we studied, the quantum effects appear not to be of
great importance. This is encouragement that, when attempts to go
beyond the spherical symmetry assumed in this work, the simpler
classical approach may suffice to reveal most of the important physics.

In all of the work in this paper (and in most of the literature), we
have assumed spherical symmetry of the solutions. Regrettably, this
geometry is too simple for many realistic applications. The intrinsic
sources are reasonably uniform in lego variables, not spherical
coordinates, and this geometry needs more detailed study. In addition,
fluctuations about the mean behavior are very important. A piece of DCC
with relatively large transverse velocity will look in the laboratory
like a coreless minijet, with contents containing small relative
momenta. So the source distributions most relevant to DCC searches in
high energy hadron collisions should not only be described in lego
variables, but also contain minijet clusterings.

However, in defense of what we do, each piece of DCC in momentum space
is a cluster of pions of near identical momenta --- a ``snowball'' ---
which has a local rest frame \cite{bj,bjminn}. In a snowball
rest-frame, the calculations we make should be a reasonable description
of the dynamics of that particular snowball. But one needs to know how
the chiral orientations and probability of occurrence of snowballs
which are neighboring in momentum-space are correlated. Very little
work on this exists.

In addition, one should average over sources more broadly. This
includes not only the properties of the intrinsic sources discussed
above, but also the initial conditions imposed on the chiral fields at
early proper time, {\it i.e.} at the onset of the chiral symmetry
breaking. A good starting point will be to do this for the classical
version of the interesting model of Lampert, Dawson, and Cooper
\cite{lampert}, truncated at large times as described in this paper.

The final product of all this should be a generating functional for DCC
particle production, which is an average over sources of a Gaussian
generating functional characteristic of a coherent-state and classical
solution produced by a specific source (see, for example, reference
\cite{apw} for the formalism). However, even when this is attained, it
still leaves open the more difficult problem of synthesizing such a
generating functional with one for generic production, since there is
not yet a consensus on what represents a good choice for the latter.

The emphasis we make in this paper on DCC sources makes the formalism
look more and more similar to what is used to describe Bose-Einstein
correlations \cite{apw}. There is certainly a close relationship
\cite{AndrOther}. What we believe special about the baked-alaska
scenario, even accepting only the broad outlines, is that there is
assumed to be nontrivial dynamics occurring deep within the future
light cone. Irrespective of details of our modeling, the presence of
such dynamics would be a new element in the description of
hadron-hadron collisions at high energies.

\section*{Acknowledgements}
One of us (JDB) gratefully acknowledges the contributions of Marvin
Weinstein, Cyrus Taylor, and Kenneth Kowalski in collaborative work on
these lines which began long ago but never reached fruition. He also
thanks A. Anselm for useful discussions, and all his collaborators in
the Minimax test/experiment T864 at Fermilab for much support and
stimulation. One of us (GA-C) gratefully acknowledges useful
discussions with Mike Birse, Abdelatif Abada, and Melissa Lampert. We
also thank Melissa Lampert for providing data points from the analysis
in Ref.~\cite{lampert}. In addition we thank the participants in last
year's Trento DCC workshop, organized by Rob Pisarski and Jorgen
Randrup, for much criticism and stimulation. This work was supported in
part by funds provided by the Foundation Blanceflor
Boncompagni-Ludovisi, P.P.A.R.C., U.S. Department of Energy contract
DE-AC03-76SF00515, the American Trust for Oxford University (George
Eastman Visiting Professorship), CSN and KV (Sweden), ORS and OOB
(Oxford), and the Sir Richard Stapley Educational Trust (Kent).

\baselineskip 12pt plus .5pt minus .5pt

\end{document}